\documentstyle[11pt,a4]{article}

\renewcommand{\epsilon}{\varepsilon}
\begin{document}
\title{Thermodynamic equilibrium in the expanding universe}

\author{Winfried Zimdahl\footnote{Fakult\"at f\"ur Physik, Universit\"at Konstanz,
PF 5560 M678, 
D-78434  Konstanz, Germany, electronic address: winfried.zimdahl@uni-konstanz.de} 
and Alexander B. Balakin \footnote {Department of General Relativity and Gravitation,  
Kazan State University, 420008 Kazan, Russia, electronic address: dulkyn@ant.ksc.iasnet.ru}
}

\date{\today}
\maketitle
\begin{abstract}
We show that a relativistic gas may be at ``global'' equilibrium in the expanding universe for any equation of state  
$0 < p \leq \rho /3$, provided that the gas particles move under the influence of a self-interacting, effective one-particle force in between elastic binary collisions. 
In the force-free limit we recover the equilibrium conditions for ultrarelativistic matter which imply the existence of a conformal timelike Killing vector. 
\end{abstract}

\section{Introduction}
The Lie derivative $\pounds _{\xi _{i}}g _{ab}$ of the metric  
$g _{ab}$ with respect to  $\xi _{i}$ may be used to characterize spacetime symmetries. The vector $\xi ^{i}$ is a Killing vector (KV) if 
$\pounds _{\xi _{i}}g _{ab}={\rm 0}$ is satisfied. 
In fluid spacetimes the vector $\xi ^{i}=u ^{i}/T$, where $u ^{i}$ is the 
four-velocity of the fluid, normalized according to $u ^{i}u _{i} = - 1$, and $T$ is its equilibrium temperature, is of particular interest. 
In case $u ^{i}/T$ is a KV, the corresponding spacetime is stationary. 
In a cosmological context one may consider the possibility of $u ^{i}/T$ to be a conformal Killing vector (CKV): 
\begin{equation}
\pounds _{_{\frac{u _{a}}{T}}} g _{ik} \equiv   \left(\frac{u _{i}}{T} \right)_{;k} 
+ \left(\frac{u _{k}}{T} \right)_{;i} = {\rm 2} \phi  g_{ik}  \ . 
\label{1}
\end{equation}
It is easy to realize that the latter relation implies 
\begin{equation}
\phi =  \frac{1}{3}
\frac{\Theta }{T}\ ,
\label{2}
\end{equation}
as well as 
\begin{equation}
\frac{T}{2}u ^{i}u ^{k}\pounds _{_{\frac{u _{a}}{T}}}g _{ik} 
\equiv   \frac{\dot{T}}{T} = - {\rm \frac{\Theta }{3}} 
\equiv  - \frac{\dot{a}}{a}\ , 
\label{3}
\end{equation}
where $\dot{T} \equiv  T _{,a}u ^{a}$ and $\Theta \equiv  u ^{i}_{;i}$ is the fluid expansion. 
The quantity $a$ is a length scale which in the special case of a homogeneous and isotropic universe coincides with the scale factor of the Robertson Walker metric. 
Equation (\ref{3}) implies 
$T \propto a ^{-1}$ which describes the temperature behaviour of an ultrarelativistic fluid (radiation) in the expanding universe. 
The ultrarelativistic limit is clearly singled out by the CKV property, although no equation of state was used to obtain 
the temperature behaviour (\ref{3}). 
It is well known too, that for a relativistic Boltzmann gas the CKV property of $u ^{a}/T$ follows as an equilibrium condition (``global'' equilibrium) from Boltzmann's equation in the corresponding limit of massless particles (radiation) \cite{Stew,Ehl,Groot,Neugeb}. 
For any other equation of state the ``global'' equilibrium condition is 
$\pounds _{u _{a}/T}g _{ik}={\rm 0}$, a requirement which cannot be realized in an expanding universe. 
The ultrarelativistic equation of state is a singular case both macroscopically and microscopically. 
This circumstance may give rise to the question as to whether there exist symmetries which are compatible with a nonvanishing expansion and which prefer different equations of state in a similar sense in which the CKV symmetry prefers an ultrarelativistic equation of state. 
It is this question which we will address and answer affirmatively in the present paper. 

Based on the observation that the temperature law (\ref{3}), which is implied by the CKV property (\ref{1}) of $u _{i}/T$, may independently be derived from the thermodynamics of an ultrarelativistic fluid, we find, in section 2, an expression for $\pounds _{u _{a}/T}g _{ik}$ which fits the temperature law for a fluid with arbitrary equations of state. 
Section 3 shows how this expression follows from the Boltzmann equation as an equilibrium condition for the one-particle distribution function for a relativistic gas in a force field. 
In section 4 we consider the microscopic particle motion in this force field and confirm explicitly that the equilibrium distribution of the gas particles is maintained as the universe expands. 
Units have been chosen so that
$c = k_{B} = 1$.

\section{Spacetime symmetries}
In a first step we recall how the temperature law (\ref{3}), which we obtained here as a consequence of the CKV property (\ref{1}), is alternatively and independently derived on a thermodynamical basis. 
Let us consider a perfect fluid with the energy-momentum tensor 
\begin{equation}
T ^{ik} = \rho  u ^{i}u ^{k} + p h ^{ik}\ ,
\label{4}
\end{equation}
where $\rho $ is the energy density measured by a comoving observer, $p$ is the equilibrium pressure, 
and $h ^{ik}$ is the projection tensor $h ^{ik}=g ^{ik}+u ^{i}u ^{k}$. 
Local energy-momentum conservation $T ^{ik}_{\ ;k} = 0$ implies 
\begin{equation}
- u _{i}T ^{ik}_{\ ;k} = \dot{\rho } + \Theta \left(\rho + p \right) = 0
\label{5}
\end{equation}
and 
\begin{equation}
h _{im}T ^{ik}_{\ ;k} = 
\left(\rho + p\right)\dot{u} _{m} 
+ \nabla  _{m}p  = 0 \ .
\label{6}
\end{equation}
The number $N$ of particles in a comoving volume $a ^{3}$ is $N = na ^{3}$, 
where $n$ is the particle number density. 
The corresponding particle number flow vector is $N ^{a} = n u ^{a}$ and the particle number conservation law may be written as 
\begin{equation}
N ^{a}_{;a} = \dot{n} + \Theta n = 0 \ .
\label{7}
\end{equation}
Given equations of state in the general form 
\begin{equation}
p = p \left(n,T \right)\ ,\ \ \ \ \ \ \ \ \ 
\rho = \rho \left(n,T \right)\ ,
\label{8}
\end{equation}
differentiation of the latter relation and using 
the balances (\ref{5}) and  (\ref{7}) 
yields 
\begin{equation}
\frac{\dot{T}}{T} = - \Theta 
\frac{\partial p}{\partial \rho } \ , 
\label{9}
\end{equation}
where the abbreviation
\[
\frac{\partial{p}}{\partial{\rho }} \equiv  
\frac{\left(\partial p/ \partial T \right)_{n}}
{\left(\partial \rho / \partial T \right)_{n}} \ ,
\]
as well as the general relation
\[
\left(\frac{\partial{\rho }}{\partial{n}} \right)_{T} = \frac{\rho + p}{n} 
- \frac{T}{n}\left(\frac{\partial{p}}{\partial{T}} \right)_{n}
\] 
have been used. 
For an ultrarelativisic equation of state  
$\partial p/ \partial \rho = 1/3$ we recover the temperature law (\ref{3}). 
It is only this specific equation of state for which  the expressions 
(\ref{9}) and (\ref{3}) are compatible. 

It is now tempting, however, to combine the temperature behaviour (\ref{9}) with the Lie derivative of the metric also for a general equation of state. 
It is easy to realize that for a choice 
\begin{equation}
\pounds _{_{\frac{u _{a}}{T}}} g _{ik}  = 
{\rm 2} \phi \left[g _{ik} - 
\left({\rm 3} \frac{\partial{p}}{\partial{\rho }} - {\rm 1}\right)u _{i}u _{k} \right] \ ,
\label{10}
\end{equation}
where $\phi $ is again given by (\ref{2}), the temperature law (\ref{9}) is reproduced for any equation of state. 
In the ultrarelativistic limit $\partial p/ \partial \rho = 1/3$ 
we recover the CKV case (\ref{3}). 
The expression (\ref{10}) for the Lie derivative makes the fluid spacetime symmetry depend on the fluid equation of state. 
The CKV property (\ref{3}) is no longer a singular case. 

For a gas, the KV and CKV conditions are known to follow from Boltzmann's equation  as equilibrium conditions for the one-particle distribution function \cite{Stew,Ehl,Groot,Neugeb}. In the following section we will clarify the conditions under which the general expression (\ref{10}) is related to equilibrium properties of the gas. 
The essential point will be that in order to obtain (\ref{10}) as equilibrium condition for the one-paricle distribution function, the gas particles will have to move in a suitable force field. 
The ultrarelativistic case, corresponding to the CKV property, will then follow in the limit of a vanishing force. 

\section{Kinetic theory in a force field}
Let the particles of a gas in between elastic binary collisions move 
under the influence of a four force 
$F ^{i}=F ^{i}\left(x,p \right)$ 
according to the equations of motion 
\begin{equation}
\frac{\mbox{d} x ^{i}}{\mbox{d} \gamma  } = p ^{i}\ ,
\ \ \ \ \ \ 
\frac{\mbox{D} p ^{i}}{\mbox{d} \gamma  } = m F ^{i}\ , 
\label{11}
\end{equation} 
where $\gamma  $ is a parameter along their worldline which for massive particles 
may be related to the proper time $\tau $ by $\gamma   = \tau /m$. 
Since the particle four-momenta are normalized according to 
$p ^{i}p _{i} = - m ^{2}$, the force $F ^{i}$ has to satisfy the relation 
$p _{i}F ^{i} = 0$. 
The one-particle distribution function 
$ f = f\left(x,p\right)$ then 
obeys the Boltzmann equation (see, e.g., \cite{Stew,Ehl,Groot,Neugeb}) 
\begin{equation} 
p^{i}f,_{i} - \Gamma^{i}_{kl}p^{k}p^{l}
\frac{\partial f}{\partial
p^{i}} + m F ^{i}\frac{\partial{f}}{\partial{p ^{i}}}
 = C\left[f\right]   \mbox{ . } 
\label{12}
\end{equation} 
$C[f]$ is Boltzmann's collision integral.   
The effective one-particle force $F ^{i}$ is supposed to model different kinds of interactions in a simple manner \cite{ZTP,ZPRD98,ZiBa1,ZiBa2}. 
Its specific structure in the present context will be determined below. 

The particle number flow 4-vector 
$N^{i}$ and the energy momentum tensor $T^{ik}$ are
defined in a standard way (see, e.g., \cite{Ehl}) as 
\begin{equation}
N^{i} = \int \mbox{d}Pp^{i}f\left(x,p\right) \mbox{ , } 
\ \ \ 
T^{ik} = \int \mbox{d}P p^{i}p^{k}f\left(x,p\right) \mbox{ .} 
\label{13}
\end{equation}
The integrals in the definitions (\ref{13}) and in the following  
are integrals over the entire mass shell 
$p^{i}p_{i} = - m^{2}$. 
The entropy flow vector $S^{a}$ is given by \cite{Ehl,IS} 
\begin{equation}
S^{a} = - \int p^{a}\left[
f\ln f - f\right]\mbox{d}P \mbox{ , }
\label{14}
\end{equation}
where we have restricted ourselves to the case of 
classical Maxwell-Boltzmann particles. 

Using well-known general relations (see, e.g., \cite{Stew})  
we find 
\begin{eqnarray}
N^{a}_{;a} &=& \int \left(C\left[f\right] 
- m F ^{i}\frac{\partial{f}}{\partial{p ^{i}}}\right) \mbox{d}P 
\mbox{ , } \ \ \nonumber\\
T^{ak}_{\ ;k} &=&  \int p^{a}\left(C\left[f\right] 
- m F ^{i}\frac{\partial{f}}{\partial{p ^{i}}}\right) 
\mbox{d}P
\mbox{ , } 
\label{15}
\end{eqnarray}
and 
\begin{equation}
S^{a}_{;a} = - \int \ln f 
\left(C\left[f\right] 
- m F ^{i}\frac{\partial{f}}{\partial{p ^{i}}}\right) \mbox{d}P
\mbox{ .} 
\label{16}
\end{equation}
Since we are interested in the equilibrium properties of the gas we will restrict ourselves to collisional equilibrium from now on. 
Under this condition  $\ln f$ in Eq. 
(\ref{16}) 
is a linear combination of the collision invariants 
$1$ and $p^{a}$ and the collision integral $C \left[f \right]$ vanishes.  
The corresponding equilibrium distribution function 
becomes (see, e.g., \cite{Ehl}) 
\begin{equation}
f^{0}\left(x, p\right) = 
\exp{\left[\alpha + \beta_{a}p^{a}\right] } 
\mbox{ , }
\label{17}
\end{equation}
where $\alpha = \alpha\left(x\right)$ and 
$\beta_{a}\left(x \right)$ is timelike. 
Inserting the equilibrium distribution function (\ref{17}) 
into the Boltzmann equation (\ref{12}) one obtains
\begin{equation}
p^{a}\alpha_{,a} +
\beta_{\left(a;b\right)}p^{a}p^{b}   
=  - m \beta _{i}F ^{i} 
\mbox{ .} 
\label{18}
\end{equation} 
For a vanishing force $F ^{i}$ the latter condition reduces to the  ``global'' equilibrium condition of standard relativistic kinetic theory, i.e., 
to $\alpha = {\rm const}$ and either to the KV condition  for $m > 0$, or to the CKV condition (\ref{1}) for $m = 0$. 
Use of the equilibrium distribution function (\ref{17}) in the balances (\ref{15}) yields 
\begin{eqnarray}
N^{a}_{;a}&=&-m \beta _{i}\int F ^{i}f ^{0}\mbox{d}P \ , \nonumber\\ 
T^{ak}_{\ ;k}&=&-m \beta _{i}\int p^{a}F ^{i}f ^{0}\mbox{d}P \ .
\label{19}
\end{eqnarray}
For the entropy production density (\ref{16}) we find 
\begin{equation}
S^{a}_{;a} = m \beta _{i} \int \left[\alpha + \beta _{a}p ^{a} \right]
F ^{i}f ^{0}
\mbox{d}P 
=  - \alpha N^{a}_{;a} 
- \beta_{a}T^{ab}_{\ ;b}
\mbox{ . }
\label{20}
\end{equation}  

With $f$ replaced by $f^{0}$ in the definitions 
(\ref{13}) and (\ref{14}), $N^{a}$, $T^{ab}$ and $S^{a}$ may be 
split with respect to the unique 4-velocity $u^{a}$ according to 
\begin{equation}
N^{a} = nu^{a} \mbox{ , \ \ }
T^{ab} = \rho u^{a}u^{b} + p h^{ab} \mbox{ , \ \ }
S^{a} = nsu^{a} \mbox{  , }
\label{21}
\end{equation}
where $u ^{a}$, $h ^{ab}$,  
$n$,  $\rho$, and $p$ 
may be identified with the corresponding quantities of the previous section.  
$s$ is the entropy per particle. 
The exact integral expressions for $n$, $\rho$, $p$ are given by (see, e.g., \cite{Groot}), 
\begin{equation}
n =  \frac{p}{T} = \frac{4\pi m^{2}T}
{\left(2\pi\right)^{3}}K_{2}\left(
\frac{m}{T}\right) \exp{\left[\alpha \right]}\ ,
\label{22}
\end{equation}
and  
\begin{equation}
e =  \frac{\rho }{n} = m\frac{K_{1}\left(\frac{m}{T}\right)}
{K_{2}\left(\frac{m}{T}\right)} + 3 T = 
m\frac{K_{3}\left(\frac{m}{T}\right)}
{K_{2}\left(\frac{m}{T}\right)} -   T \ .
\label{23}
\end{equation}
The quantities $K_{n}$ are modified Bessel functions of the second
kind \cite{Groot}.  
The entropy per particle $s$ is 
\begin{equation}
s = \frac{\rho + p}{nT} - \alpha \ .
\label{24}
\end{equation}

In order to evaluate the equilibrium condition (\ref{18}), we expand the quantity $\beta _{i}F ^{i}$ in a power series in 
$p ^{a}$. According to the structure of the left-hand side of 
(\ref{18}), it will be sufficient to consider terms up to second order in the momentum: 
\begin{equation}
\beta _{i}F ^{i} = \beta _{i}F ^{i}_{a}p ^{a} 
+ \beta _{i}F ^{i}_{kl}p ^{k}p ^{l}  \ .
\label{25}
\end{equation}
The quantities 
$F ^{i}_{a}$ and  $F ^{i}_{kl}$  
are spacetime functions to be determined below, i.e., they do not depend on $p ^{a}$. 
We emphasize that it is only the projection $\beta _{i}F ^{i}$ of 
$F ^{i}$ which is expanded in a power series, not the force  itself. 
The orthogonality of the latter to $p ^{i}$ may always be guaranteed 
\cite{ZiBa1}. 

With the expansion (\ref{25}) 
the equilibrium condition (\ref{18}) splits into  
\begin{equation}
\alpha _{,a} = - m \beta _{i}F _{a}^{i}
\label{26}
\end{equation}
and 
\begin{equation}
\beta _{\left(k;l \right)} = - m \beta _{i}F ^{i}_{kl}\ .
\label{27}
\end{equation}
With the identification $\beta _{i}=u _{i}/T$ it is obvious that the choice 
\begin{equation}
- m \beta _{i}F ^{i}_{ab}  = \frac{\Theta }{3T}
\left[g _{ab} - 
\left(3 \frac{\partial{p}}{\partial{\rho }} - 1\right)u _{a}u _{b} \right] \ 
\label{28}
\end{equation} 
just reproduces (\ref{10}). 
{\it The symmetry requirement (\ref{10}) is recovered as an equilibrium condition for the one-particle distribution function for gas particles in a specific force field.} 
For $\partial p/ \partial \rho = 1/3$ the condition (\ref{27}) with (\ref{28}) reduces to the CKV condition (\ref{1}). 

The condition (\ref{26}) may be written as  
\begin{equation}
- m \beta _{i}F ^{i}_{a} = - \dot{\alpha}u _{a} + \nabla  _{a} \alpha \ ,
\label{29}
\end{equation}
where $\nabla  _{a}\alpha \equiv h _{a}^{b}\alpha _{,b} $.  
Applying the ansatz (\ref{25}) in the balances (\ref{19}), we find  \begin{equation}
N ^{a}_{;a} = - m \beta _{i}\left(F ^{i}_{a}N ^{a} 
+ F ^{i}_{ab}T ^{ab} \right)\ ,
\label{30}
\end{equation}
and 
\begin{equation}
T ^{ak}_{\ ;k} = - m \beta _{i}\left(F ^{i}_{b}T ^{ab} 
+ F ^{i}_{kl}M^{akl} \right)\ ,
\label{31}
\end{equation}
where $M ^{akl} = \int \mbox{d}P f ^{0}p ^{a}p ^{k}p ^{l}$ is the third moment of the equilibrium distribution function. 
Use of the expressions (\ref{28}) and (\ref{29}) allows us to write 
the particle number balance (\ref{30}) as  
\begin{equation}
N ^{a}_{;a} = n \dot{\alpha} + n \Theta 
\left[1 - \frac{\rho }{nT}\frac{\partial{p}}{\partial{\rho }} \right]\ .
\label{32}
\end{equation}
With the identification $\alpha = \mu /T$, the quantity $\dot{\alpha}$ may be calculated from the Gibbs-Duhem relation 
\begin{equation}
\mbox{d} p = \left(\rho + p \right)\frac{\mbox{d} T}{T} 
+ n T \mbox{d} \left(\frac{\mu }{T} \right)\ .
\label{33}
\end{equation}
With $p = n T$ and the temperature law (\ref{9}) we obtain 
\begin{equation}
\dot{\alpha} = \left(\frac{\mu }{T} \right)^{\displaystyle \cdot} = 
- \Theta 
\left[1 - \frac{\rho }{nT}\frac{\partial{p}}{\partial{\rho }} \right]\ .
\label{34}
\end{equation} 
Inserting this result into (\ref{32}) 
we recover formula (\ref{7}) 
i.e., the particle number is indeed conserved. 
Analogously, we find from Eq. (\ref{31}),  
\begin{equation}
u _{a}T ^{ak}_{\  ;k} = - \dot{\alpha}\rho 
+ \phi u _{a} \left[g _{kl} - \left(3 \frac{\partial{p}}{\partial{\rho }} - 1\right) u _{k}u_{l}\right]M ^{akl}\ .
\label{35}
\end{equation}
Obviously, one has  
\begin{equation}
g _{kl}M ^{akl} = g _{kl}\int \mbox{d}P f ^{0}p ^{a}p ^{k}p ^{l} 
= - n m ^{2}u ^{a}\ .
\label{36}
\end{equation}
The expression $u _{a}u _{k}u_{l}M ^{akl}$ in the balance (\ref{35}) may be evaluated by use of the relations 
\begin{equation}
u _{a}p ^{a}f ^{0} = \frac{\partial{f ^{0}}}
{\partial \left({\frac{1}{T}} \right)}\ ,
\label{37}
\end{equation}
(the derivative has to be taken for $\alpha = const$), 
\begin{equation}
\frac{\partial{p}}{\partial{\frac{1}{T}}} = - T \left(\rho + p \right) 
\mbox{\ \ \ \ }
{\rm and}
\mbox{\ \ \ }
\frac{\partial{\rho }}{\partial{\frac{1}{T}}} 
= - 3T \left(\rho + p \right) - pT \left(\frac{m}{T} \right)^{2}\ ,
\label{38}
\end{equation}
which follow with the help of the properties 
\begin{equation}
\frac{\mbox{d}}{\mbox{d}z}\left(\frac{K _{n}\left(z \right)}{z ^{n}} \right) 
= - \frac{K _{n+1}\left(z \right)}{z ^{n}} 
\mbox{\ \ \ \ }
{\rm and}
\mbox{\ \ \ }
K _{n+1} = 2n \frac{K _{n}}{z} + K _{n-1}
\label{39}
\end{equation}
of the functions $K _{n}$ \cite{Groot}. 
The result is 
\begin{equation}
u _{a}u _{k}u_{l}M ^{akl} = - 3T \left(\rho + p \right) 
- p T \left(\frac{m}{T} \right)^{2}\ .
\label{40}
\end{equation} 
The energy balance (\ref{35}) then becomes  
\begin{equation}
u _{a}T ^{ak}_{\  ;k} = n T \Theta \frac{\partial{p}}{\partial{\rho }}
\left[\left(\frac{m}{T} \right)^{2} - 1 + 5 \frac{\rho + p}{nT} 
- \left(\frac{\rho + p}{nT} \right)^{2}\right] - nT \Theta \ .
\label{41}
\end{equation} 
Since the expression in the bracket of the last relation is \cite{ZTP}
\begin{equation}
\left(\frac{m}{T} \right)^{2} - 1 + 5 \frac{\rho + p}{nT} 
- \left(\frac{\rho + p}{nT} \right)^{2} = 
\frac{\partial{\rho }}{\partial{p}}\ ,
\label{42}
\end{equation} 
one recovers the local energy conservation (\ref{5}). 
Similarly one shows that the momentum balance (\ref{6}) is fulfilled identically, which implies 
\[
\nabla _{a}\alpha = - \left(\dot{u}_{a} + \frac{\nabla  _{a}T}{T} \right)
\frac{\rho + p}{nT} = 0 \ .
\] 
Thus we have clarified that the self-interacting force term (\ref{25}) 
with (\ref{28}) and (\ref{29}) both implies the symmetry condition (\ref{10}) 
and  at the same time respects the conservation laws for particle number and energy momentum. As a consequence, the entropy production density (\ref{20}) vanishes also. 
We emphasize that the mentioned conservation properties single out a very specific self-interaction. 
For a general force term of of the type (\ref{25}) with a coefficient different from (\ref{28}) none of the moments of the distribution function (\ref{17}) will be conserved. 
This has been explicitly demonstrated for two cases, the first of which is a conformal symmetry (\ref{1}) for massive particles \cite{ZTP,ZPRD98,ZiBa1}. 
While such a configuration does not exist for a conserved particle number, it turned out to be possible if these particles were created at a specific rate. 
The second case was based on a modification of the conformal 
symmetry (\ref{1})  according to 
$\pounds _{u _{a}/T}g _{ik} =h _{ik}$ \cite{ZiBa2}. 
The corresponding self-interaction implies particle production at a rate which coincides with the expansion rate of the universe.. 
The source terms in the balances for the second moments in each of these cases were mapped onto bulk pressures of effective energy momentum tensors which are conserved. 
Although self-interactions of this kind are entropy producing, the corresponding states are ``generalized'' equilibrium states \cite{ZPRD98} since the gas particles are still characterized by the equilibrium distribution function (\ref{17}). 
The case dealt with here is singled out by the fact that the ``generalized'' equilibrium coincides with the conventional ``global'' equilibrium with conserved moments of the distribution function. 

\section{Microscopic particle motion } 
Knowledge of the coefficients (\ref{28}) and (\ref{29}) 
in (\ref{25})  allows us to study 
the motion of the gas particles explicitly. 
Contracting the equation of motion in (\ref{11}) with the macroscopic four-velocity results in 
\begin{equation}
\frac{\mbox{D}\left(u _{i}p ^{i} \right)}{\mbox{d}\tau } 
= u _{i}F ^{i} + \frac{1}{m}u _{i;k}p ^{i}p ^{k} \ ,
\label{43}
\end{equation}
where we have used that 
\[
\frac{\mbox{D} u ^{i}}{\mbox{d} \tau } = u ^{i}_{;n}\frac{p ^{n}}{m}\ .
\]
With 
the well-known decomposition of the covariant derivative of the 
four-velocity \cite{Ehlers,Ellis}, 
\begin{equation}
u _{i;n} = - \dot{u}_{i}u _{n} + \sigma _{in} +  \omega _{in}  
+ \frac{\Theta }{3}h _{in}\ ,
\label{44}
\end{equation}
where 
$\sigma_{ab} = h_{a}^{c}h_{b}^{d}u_{\left(c;d\right)}-h _{ab}\Theta /3$ and 
$\omega_{ab} = h_{a}^{c}h_{b}^{d}u_{\left[c;d\right]}$, we find generally 
\begin{equation}
\frac{\mbox{D}\left(u _{i}p ^{i} \right)}{\mbox{d}\tau } 
= u _{i}F ^{i} + \frac{1}{3m} \Theta h _{ik}p ^{i}p ^{k} 
+ \frac{1}{m}\sigma _{ik}p ^{i}p ^{k} 
- \frac{1}{m}\dot{u}_{i}u _{k}p ^{i}p ^{k}\ .
\label{45}
\end{equation} 
In the following we investigate this equation for the homogeneous case 
$\dot{u}_{i}= 0$. 
(Notice that $\sigma _{ab} = \dot{u}_{a} + \nabla  _{a}T /T = 0$ are consequences of (\ref{10})). 
For $u _{i}F ^{i}$ we have under such conditions 
\begin{equation}
u _{i}F ^{i} = \frac{T}{m} \left[- \dot{\alpha}E 
+ m ^{2}\frac{\Theta }{3T} 
+ \frac{\Theta }{3T}E ^{2}\left(3 \frac{\partial{p}}{\partial{\rho }} - 1\right)\right]\ ,
\label{46}
\end{equation}
where $E \equiv - u _{i}p ^{i}$. 
Equation (\ref{45}) simplifies to 
\begin{equation}
\frac{\mbox{d}E}{\mbox{d}\tau } = - u _{i}F ^{i} 
- \frac{\Theta }{3m}  \left(E ^{2} - m ^{2} \right)\ .
\label{47}
\end{equation}

With $d \tau = d t \left(m/E \right)$ and $d E/d t \equiv  \dot{E}$, the last equation is equivalent to 
\begin{equation}
\frac{\left(E ^{2} - m ^{2} \right)^{\displaystyle \cdot}}
{E ^{2} - m ^{2}} + \frac{\left(a ^{2} \right)^{\displaystyle \cdot}}
{a ^{2}} = -\frac{2m}{E ^{2} - m ^{2}}u _{i}F ^{i} \ .
\label{48}
\end{equation}
In the force-free case $F ^{i} = 0$, i.e., for geodesic particle motion, we find 
\begin{equation}
E ^{2} - m ^{2}  \propto a ^{-2}\ 
\mbox{\ \ \ \ \ \ }
\left(F ^{i} = 0 \right)\ ,
\label{49}
\end{equation}
implying the expected behavior $E \propto a ^{-1}$ for massless 
particles (photons), while the nonrelativistic energy $\epsilon \equiv  E - m$ with $\epsilon \ll m$ of massive particles decays as $\epsilon \propto a ^{-2}$. 

In the presence of the force term (\ref{46}), use of (\ref{34}) and (\ref{9}) allows us to obtain 
\begin{equation}
\dot{E} = \dot{\mu } + \left(E - \mu  \right)\frac{\dot{T}}{T}\ ,
\label{50}
\end{equation}
equivalent to 
\begin{equation}
- \frac{E - \mu }{T} =  \frac{\mu + u _{a}p ^{a}}{T} 
= \alpha + \beta _{a}p ^{a}  = {\rm const}\ . 
\label{51}
\end{equation}
This demonstrates explicitly that the equilibrium distribution (\ref{17}) is indeed maintained in the expanding universe for arbitrary equations of state. 

\section{Conclusion}
We have clarified the conditions under which the particles of an expanding gaseous fluid are governed by an one-particle equilibrium distribution function which satisfies the relativistic Boltzmann equation. 
In general, such kind of equilibrium is only possible if there exist suitable forces inside the system which make the particles move on nongeodesic trajectories in between equilibrium establishing elastic binary collisions described by Boltzmann's collision integral. 
These forces depend on the microscopic particle momenta as well as on macroscopic fluid variables including the equation of state. They represent a (nongravitational) self-interaction of the gas which is consistent with the conservation of particle number and energy momentum. 
In other words, the gas as a whole must mobilize specific internal forces to maintain an equilibrium distribution of the particles as the universe expands. 
The familiar  ultrarelativistic case in which  $u _{i}/T$ is a CKV corresponds to the force free limit. 
\ \\
\ \\
{\bf Acknowledgment}\\
This paper was supported by the Deutsche Forschungsgemeinschaft.


\begin{thebibliography}{99}
\bibitem{Stew} J.M. Stewart, {\it Non-equilibrium Relativistic  
Kinetic Theory}
(Springer, New York, 1971).
\bibitem{Ehl} J. Ehlers,  in {\it
General Relativity and Cosmology} ed. by B.K. Sachs
(Academic Press, New York, 1971).
\bibitem{Groot} S.R. de Groot, W.A. van Leeuwen, and Ch G. van  
Weert, {\it
Relativistic Kinetic Theory} (North Holland, Amsterdam, 1980).
\bibitem{Neugeb} G. Neugebauer, 
{\it Relativistische Thermodynamik}
(Vieweg, Braunschweig, 1980).
\bibitem{IS} W. Israel and J.M Stewart,
Ann.Phys. {\bf 118}, 341 (1979).
\bibitem{ZTP} W. Zimdahl, J. Triginer, 
and D. Pav\'{o}n,
Phys. Rev. D {\bf 54}, 6106 (1996).
\bibitem{ZPRD98}  W. Zimdahl, 
Phys. Rev. D {\bf 57}, 2245 (1998).
\bibitem{ZiBa1} W. Zimdahl and   
A.B. Balakin, Class. Quantum Grav. {\bf 15}, 3259 (1998).
\bibitem{ZiBa2} W. Zimdahl and   
A.B. Balakin, Phys. Rev. D  {\bf 58}, 063503 (1998).
\bibitem{Ehlers} J. Ehlers, Abh. Mainz Akad. Wiss. Lit. {\bf 11}, 1 (1961).  
\bibitem{Ellis} G.F.R. Ellis,  in 
Carg\`ese Lectures in Physics, vol. VI,  ed. by E. Schatzmann 
(Gordon and Breach, New York, 1973).



\end{thebibliography}
\end{document}